\documentclass{appolb}
\usepackage{graphicx}
\usepackage{comment}
\usepackage{amssymb}
\usepackage{amsmath}
\usepackage{bm}
\usepackage[
    natbib=true,
    style=numeric,
    sorting=none
]{biblatex}
\renewbibmacro{in:}{}

\newcommand{\lr}[1]{\left\langle #1\right\rangle}
\newcommand{\llrr}[1]{\left\langle\left\langle#1\right\rangle\right\rangle}

\newcommand*{\pT}{p_{\text{T}}}

\newcommand{\rhon}[1] {\rho_{#1}}
\newcommand{\cov}[1] {\mathrm{cov}_{#1}}
\newcommand{\var}[1] {\mathrm{var}(v_{ #1 }^2)}
\newcommand{\etfcal}{\Sigma E_{\mathrm{T}}}
\newcommand{\nchrec}{N_\mathrm{ch}^\mathrm{rec}}

\usepackage[mathlines,displaymath]{lineno}

\begin{document}
% \linenumbers
\title{Flow and transverse momentum correlation in Pb+Pb and Xe+Xe collisions with ATLAS: assessing the initial condition of the QGP%
\thanks{Presented at Quark Matter 2022, Krakow, Poland\\ }%
% you can use '\\' to break lines, total 6 pages for parallel talks
}
\author{Somadutta Bhatta (on behalf of the ATLAS Collaboration) 
\address{Stony Brook University, Stony Brook, New York, USA-11790}
\\[3mm]
}
\maketitle
\begin{abstract}
One important challenge in our field is to understand the initial condition of the QGP and constrain it using sensitive experimental observables. Recent studies show that the Pearson Correlation Coefficient between ${\bm V}_n$ and event-wise mean transverse momentum $[p_{\mathrm{T}}]$, $\rho_{n}({\bm V}_n,[p_{\mathrm{T}}])$ can probe number and size of sources, nuclear deformation, volume fluctuation, and initial momentum anisotropy in initial state of heavy-ion collisions. This proceeding presents new, precision measurements of ${\bm V}_n-[p_{\mathrm{T}}]$ correlation in $^{\mathrm{129}}\mathrm{Xe}+^{\mathrm{129}}\mathrm{Xe}$ and $^{\mathrm{208}}\mathrm{Pb}+^{\mathrm{208}}\mathrm{Pb}$ collisions for harmonics $n=2$, 3, and 4 using the ATLAS detector at the LHC. The values of $\rho_{n}$ shows rich and non-monotonic dependence on centrality, $p_{T}$ and $\eta$, reflecting that different ingredients of the initial state impact different regions of the phase space. The ratio of $\rho_{2}$ between the two systems in the ultra-central region suggests that ${^\mathrm{129}}\mathrm{Xe}$ has large quadrupole deformation and with a significant triaxiality. All current models fail to describe many of the observed trends in the data.
\end{abstract}
{\let\thefootnote\relax\footnote{{Copyright 2022 CERN for the benefit of the ATLAS Collaboration. Reproduction of this article or parts of it is allowed as specified in the CC-BY-4.0 license.}}}
\section{Introduction and Measurements}
Heavy-ion collisions at LHC and RHIC produce Quark-Gluon Plasma (QGP) whose space-time evolution is well described by relativistic viscous hydrodynamics. Driven by the large pressure gradients, the QGP expands rapidly in the transverse plane, and converts the spatial anisotropy in the initial state into momentum anisotropy in the final state. The collective expansion in each event is quantified by a Fourier expansion of particle distribution in azimuth given by $\frac{dN}{d\phi} = \frac{N}{2\pi} (1+2\sum_{n=1}^\infty v_n \cos\,n(\phi-\Phi_n))$, where ${\bm V}_n$ and $\Phi_n$ represent the amplitude and phase of the $n^{\mathrm{th}}$-order azimuthal flow vector ${\bm V}_n=v_ne^{{\textrm i}n\Phi_n}$. Model calculations show that the ${\bm V}_n$ is approximately proportional to initial state eccentricity $\mathcal{E}_n$ for $n=2$ and 3, as well as for $n=4$ in central collisions~\cite{Teaney:2010vd}. In addition, the fluctuations in the size of overlap area in the initial state give rise to fluctuations in radial flow which in turn, lead to event-by-event fluctuation of the average transverse momentum ($[\pT]$). The correlated fluctuations between ${\mathcal{E}}_n$ and $R$ in the initial state is transferred to final state ${\bm V}_n$-$[\pT]$ correlations via hydrodynamic expansion. A three-particle correlator has been proposed to quantify this correlation~\cite{Bozek:2016yoj}:
\begin{align} \label{eq:rho}
\rho_n =\frac{\cov{n}}{\sqrt{\var{n}}\sqrt{c_{k}}}\;,\;\cov{n} = \llrr{v_n^2\delta \pT}\;,\;\var{n}=\lr{v_n^4}-\lr{v_n^2}^2\;,\notag \\ c_k = \llrr{\delta\pT\delta\pT}\;
\tag{1}
\end{align}
where $\delta \pT = \pT-[\pT]$ and the ``$\llrr{}$'' denotes averaging over all pairs or triplets for events with similar particle multiplicity, while the ``$\left\langle\right\rangle$'' denotes averaging over events. $\rho_n$ describes initial state correlation between size and eccentricity. Another factor determining shape and size of overlap area in initial state is ``Nuclear shape'' described by the following function~\cite{bohr}:
\begin{align}\label{eq:R}
R(\theta,\phi) = R_0\left(1+\beta [\cos \gamma Y_{2,0}+ \sin\gamma Y_{2,2}]\right)
\tag{2}
\end{align}
where $R_0$ is the nuclear radius, $Y_{l,m}$ are spherical harmonics, $\beta$, and $\gamma$ are quadrupole deformation parameters. The parameter $\beta$ is the magnitude of the deformation while the angle $\gamma$, in the range $0\leq\gamma\leq 60^{\circ}$, describes the length imbalance of the three semi-axes of the ellipsoid, also known as triaxiality. Nuclear deformation parameters $\beta$ and $\gamma$ are generally determined by low energy spectroscopic measurements. But the approximations used in such an estimate works well only for even mass nuclei. Recent studies have shown that the final state observables in heavy-ion collisions are sensitive to nuclear deformation parameters~\cite{Zhang:2021kxj,Giacalone:2021udy}. In this work, we show the strength of heavy ion collisions in providing additional handle to constrain the deformation parameters for $^{129}$Xe.

The measurement of the $\cov{n}$, $\var{n}$ and $c_k$ follows the similar procedure as detailed in Ref.~\cite{Aad:2019fgl} and uses data from the ATLAS detector~\cite{ATLAS:2008xda}. The observables in each event are averaged over events with similar multiplicity and are then combined in broader multiplicity ranges of the event ensemble to obtain statistically more precise results. The Pearson coefficient $\rhon{n}$ is then obtained via Eq.~\eqref{eq:rho}. The event averaging procedure is necessary to reduce the effects of centrality fluctuation within each event class definition.

\begin{figure}[h!]
\vspace*{-0.2cm}\centering
\includegraphics[width=1.0\linewidth]{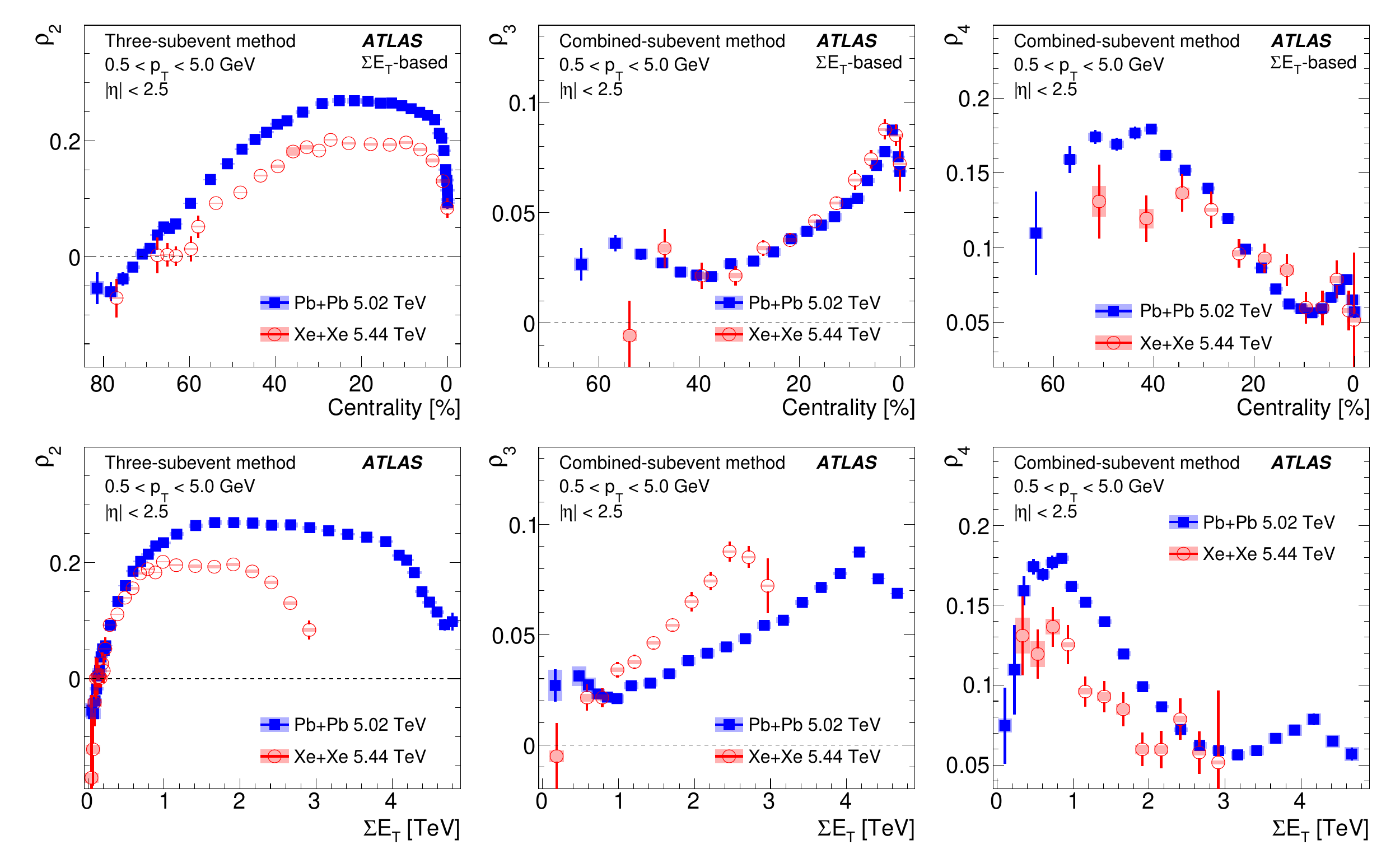}
\vspace*{-0.2cm}\caption{The centrality (top) and $\etfcal$ (bottom) dependencies of $\rhon{n}$ for $n=2$ (left), 3 (middle) and 4 (right) in Pb+Pb and Xe+Xe collisions calculated using the event-averaging procedure based on $\etfcal$~\cite{ATLAS:2022dov}.}
\label{fig:2}
\end{figure}

Figure~\ref{fig:2} provides a direct comparison of the Pb+Pb and Xe+Xe $\rhon{n}$ values as a function of centrality (top) and $\etfcal$ (bottom). The $\rhon{2}$ values reach a minimum in the peripheral collisions, increase to a positive maximum value and then decrease in the most central collisions; the $\rhon{3}$ values show a mild increase towards central collisions; the $\rhon{4}$ values show an increase then a gradual decrease towards central collisions.

In the ultracentral collision region, all the $\rhon{n}$ show a sharp decrease towards the most central collisions starting at around the location of the knee. For events having $\etfcal$ values beyond the knee all nucleons participate in the collision leading to suppression of geometrical correlations. At the same centralities, the Xe+Xe $\rhon{2}$ values are everywhere smaller than the Pb+Pb values. 

\begin{figure}[h!]
\centering
\includegraphics[width=1.0\linewidth]{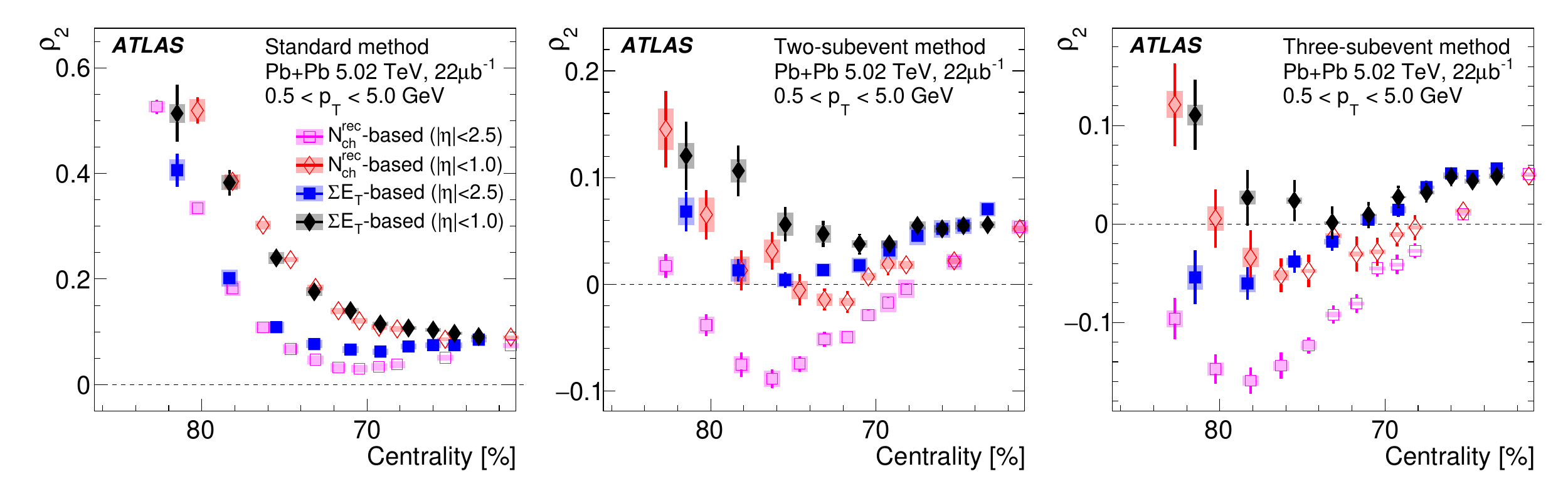}
\caption{ Centrality dependence of $\rhon{2}$ in Pb+Pb collisions in the peripheral region of 60--84\% for the standard method (left), two-subevent method (middle) and three-subevent method (right), compared between the $\nchrec$-based and $\etfcal$-based event-averaging procedures and two $\eta$ ranges~\cite{ATLAS:2022dov}.}
\label{fig:6}
\end{figure}

\begin{figure}[h!]
\centering
\includegraphics[width=1.0\linewidth]{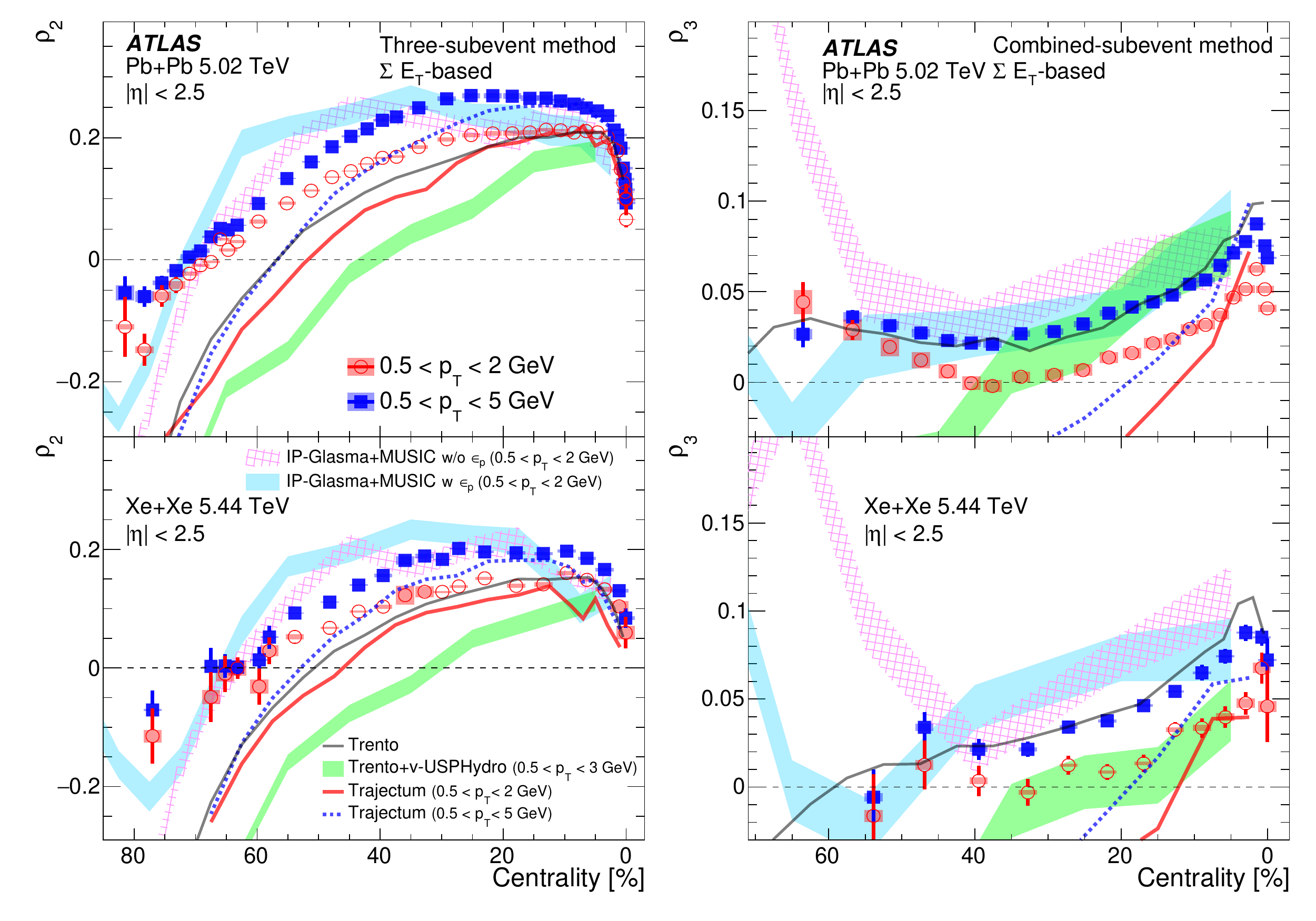}
\caption{The $\rhon{2}$ (left) and $\rhon{3}$ (right) values in Pb+Pb (top) and Xe+Xe (bottom) collisions in two $\pT$ ranges and $|\eta|<2.5$ compared with various models~\cite{ATLAS:2022dov}.}
\label{fig:8}
\end{figure}

When compared using $\etfcal$, the Pb+Pb and Xe+Xe $\rhon{2}$ values agree for small $\etfcal$ values ($\etfcal$ $<$ 0.5 TeV) but differ for larger $\etfcal$.
Recently, it was argued that $\rhon{2}$ is a sensitive probe of the nature of collectivity in small collision systems and peripheral heavy-ion collisions, in particular for isolating the contribution from initial momentum anisotropy ($\epsilon_p$) in a gluon saturation picture~\cite{Giacalone:2020byk}. The centrality dependence of $\rhon{2}$, after considering both the initial-state and final-state effects, is predicted to exhibit an increasing trend toward the most peripheral centrality. Figure~\ref{fig:6} in more detail compares the centrality dependence of $\rhon{2}$ in $|\eta|<2.5$ and $|\eta|<1$ based on $\etfcal$ and $\nchrec$ in more detail over the 60--84\% centrality range. The successive reduction of the $\rhon{2}$ from the standard method in the left panel, to the two-subevent method in the middle panel, and to the three-subevent method in the right panel is a robust feature of suppression of the non-flow correlations. The results from this measurement do not show clear evidence for initial-state momentum anisotropy.
Figure~\ref{fig:8} shows the $\rhon{2}$ and $\rhon{3}$ values for two $\pT$ ranges in Pb+Pb (top) and Xe+Xe (bottom) collisions. They are compared with initial state Trento model, 2D boost-invariant v-USPhydro model and the Trajectum model as well as full three-dimensional (3D) IP-Glasma+MUSIC hydrodynamic model.
In the 0--10\% centrality interval, where effects of nuclear deformation are important, all models show reasonable agreement with each other and with data. Trajectum model quantitatively reproduces the ordering between $0.5<\pT<2$ GeV and $0.5<\pT<5$ GeV. In noncentral collisions, these models show significant differences from each other, which were recently shown to mainly reflect the different parameter values for the initial condition such as nucleon size~\cite{Giacalone:2021clp}. In the peripheral collisions, all model predictions for $\rhon{2}$ show a sharp decrease and a sign-change, qualitatively consistent with the ATLAS data. The IP-Glasma+MUSIC model with $\epsilon_p$ shows differences from the model without $\epsilon_p$ in peripheral collisions beyond 70\% centrality.

\begin{figure}[h!]
\centering
\includegraphics[width=0.9\linewidth]{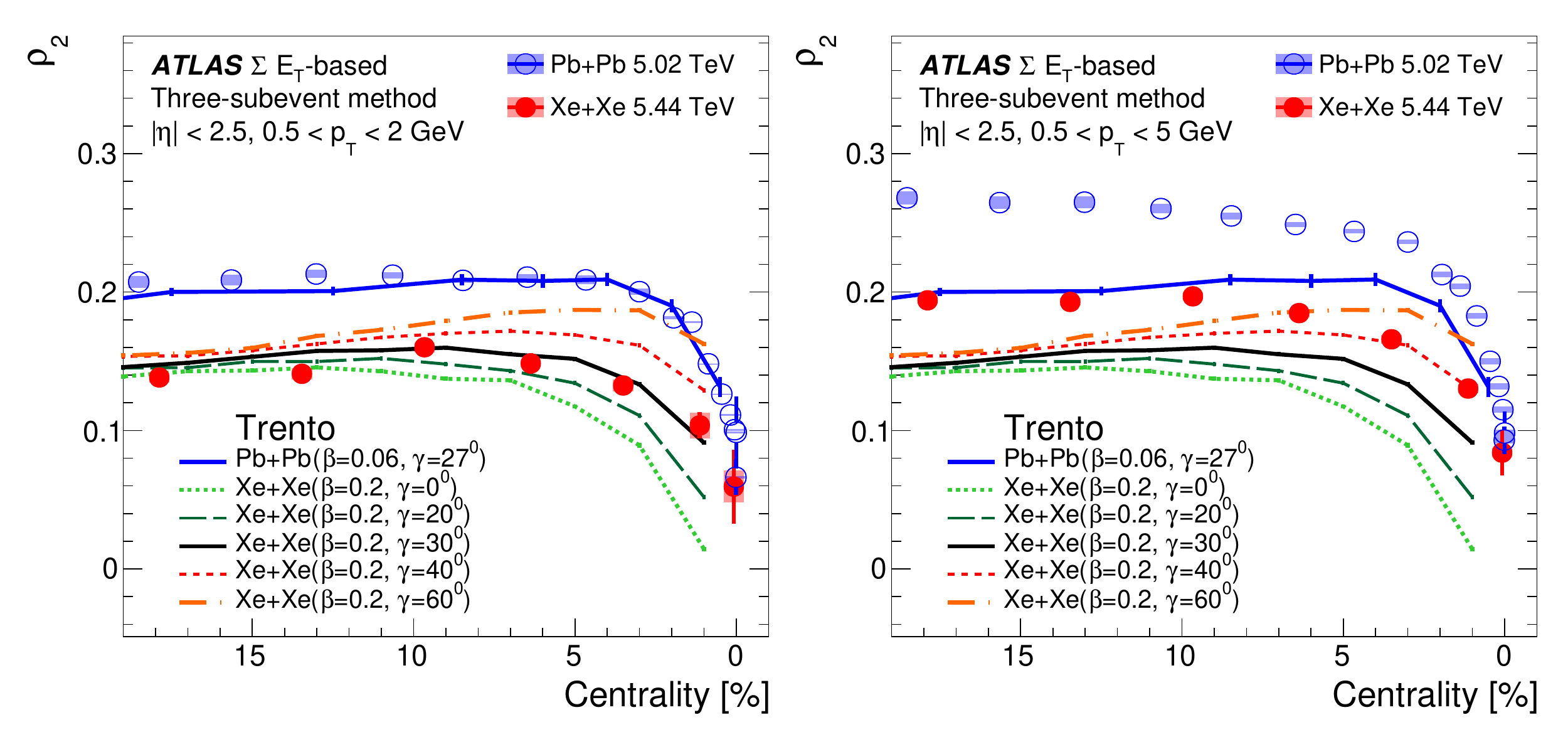}
\includegraphics[width=0.8\linewidth]{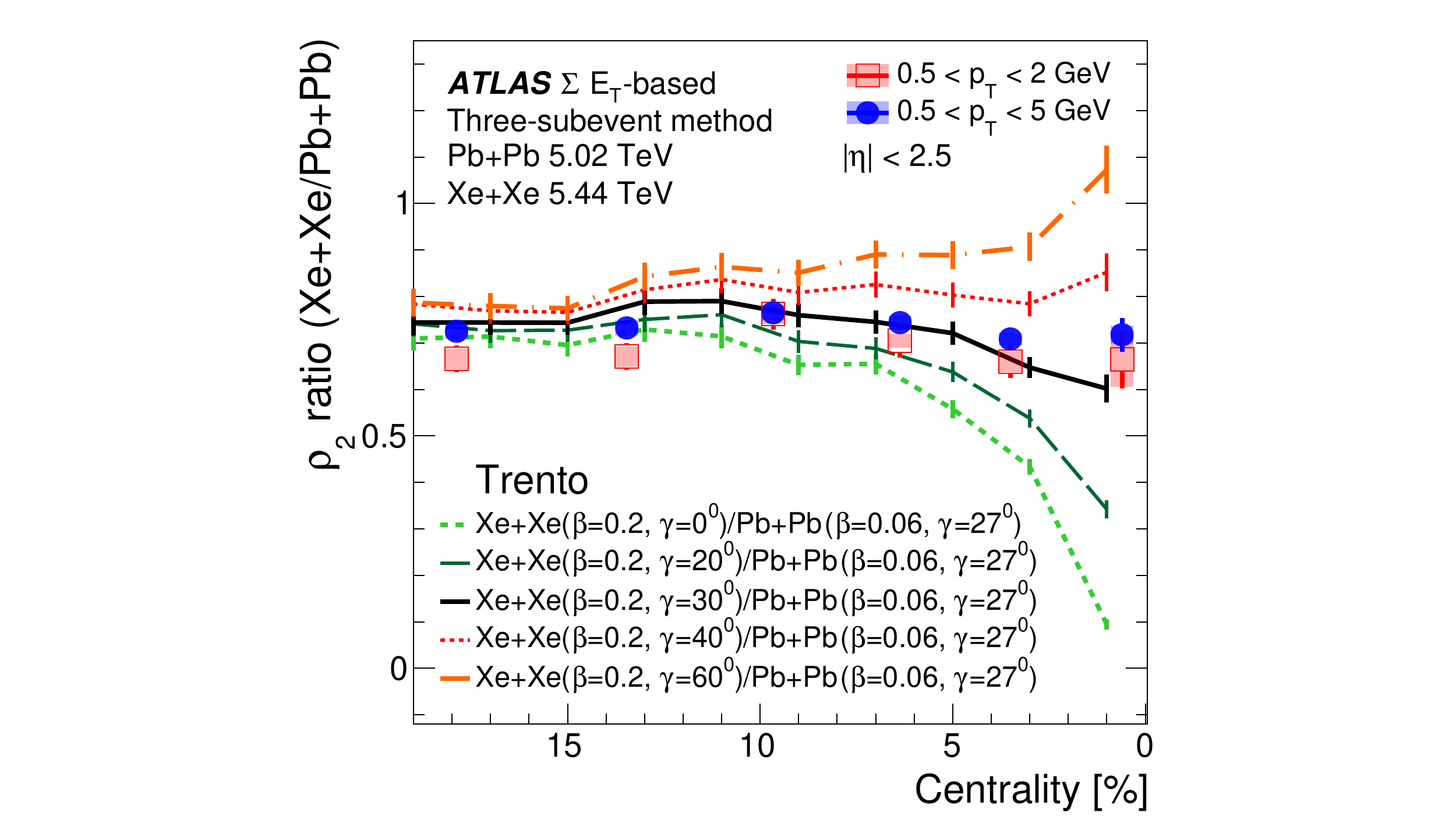}
\caption{(Top) Comparison of $\rhon{2}$ in Xe+Xe and Pb+Pb collisions with the Trento model for various $\beta$ and $\gamma$ values~\cite{Bally:2021qys} in $0.5<\pT<2$ GeV (left) and $0.5<\pT<5$ GeV (right) as a function of centrality. (Bottom) Comparison of $\rho_{2,{\mathrm{Xe+Xe}}}/\rho_{2,{\mathrm{Pb+Pb}}}$, with the Trento model for various $\beta_{\mathrm{Xe}}$ and $\gamma_{\mathrm{Xe}}$~\cite{Bally:2021qys} in two $\pT$ ranges~\cite{ATLAS:2022dov}.} 
\label{fig:9}
\end{figure}

Figure~\ref{fig:9} compares $\rhon{2}$ data in the 0--20\% centrality range with the Trento model calculations to investigate the influence of triaxiality~\cite{Bally:2021qys}. Because of the large quadrupole deformation of the $^{129}$Xe nucleus, $\beta_{\mathrm{Xe}}\sim0.2$, the $\rhon{2}$ should be sensitive to the triaxiality parameter $\gamma_{\mathrm{Xe}}$~\cite{Jia:2021qyu}. In order to cancel out the $\pT$ dependence in the data, ratios of $\rhon{2}$ values between Xe+Xe and Pb+Pb are calculated for the two $\pT$ ranges and compared with the ratios obtained in the Trento model in Figure~\ref{fig:9}. In the 0--10\% centrality range, where the predicted $\rho_2$ values show significant dependence on the triaxiality, the comparison between the model and data favors a $\gamma_{\mathrm{Xe}}\sim 30^{\circ}$. This comparison provides clear evidence that flow measurements in central heavy-ion collisions can be used to constrain the quadrupole deformation, including the triaxiality, of the colliding nuclei.
\section{Summary}
This proceeding presented experimental results on ${\bm V}_n$-$[\pT]$ correlations from ATLAS. We observed that $\rho_n$ depends on centrality fluctuations and final state $p_T$ and $\eta$ ranges but the dependencies are not very strong, suggesting they are dominated by initial state. In peripheral centralities, evidence of initial $p_T$ anisotropy (using $\rho_2$) is complicated by residual non-flow and centrality fluctuation. We also provided the first experimental constraint on triaxiality of odd mass nucleus $^{129}$Xe ($\beta$=0.2, $\gamma$=30$^{0}$) from heavy-ion collisions.
\clearpage
\printbibliography[sorting=ynt]

@article{Aad:2019fgl,
			author         = "{ATLAS Collaboration}",
			title          = "{Measurement of flow harmonics correlations with mean
												transverse momentum in lead-lead and proton-lead
												collisions at $\sqrt{s_{NN}}=5.02$ TeV with the ATLAS
												detector}",
			collaboration  = "ATLAS",
			year           = "2019",
			eprint         = "1907.05176",
			archivePrefix  = "arXiv",
			primaryClass   = "nucl-ex",
			reportNumber   = "CERN-EP-2019-130",
			SLACcitation   = "%%CITATION = ARXIV:1907.05176;%%"
}

@article{Bozek:2016yoj,
    author = "Bozek, Piotr",
    title = "{Transverse-momentum\textendash{}flow correlations in relativistic heavy-ion collisions}",
    eprint = "1601.04513",
    archivePrefix = "arXiv",
    primaryClass = "nucl-th",
    doi = "10.1103/PhysRevC.93.044908",
    journal = "Phys. Rev. C",
    volume = "93",
    number = "4",
    pages = "044908",
    year = "2016"
}

@article{Teaney:2010vd,
			author         = "Teaney, Derek and Yan, Li",
			title          = "{Triangularity and dipole asymmetry in relativistic heavy ion
												collisions}",
			journal        = "Phys.~Rev.~C",
			volume         = "83",
			pages          = "064904",
			doi            = "10.1103/PhysRevC.83.064904",
			year           = "2011",
			eprint         = "1010.1876",
			archivePrefix  = "arXiv",
			primaryClass   = "nucl-th",
			xSLACcitation   = "%%CITATION = ARXIV:1010.1876;%%",
}

@article{ATLAS:2008xda,
    author = "ATLAS Collaboration",
    collaboration = "ATLAS",
    title = "{The ATLAS Experiment at the CERN Large Hadron Collider}",
    doi = "10.1088/1748-0221/3/08/S08003",
    journal = "JINST",
    volume = "3",
    pages = "S08003",
    year = "2008"
}

@article{Giacalone:2020byk,
    author = {Giacalone, Giuliano and Schenke, Bj\"orn and Shen, Chun},
    title = "{Observable signatures of initial state momentum anisotropies in nuclear collisions}",
    eprint = "2006.15721",
    archivePrefix = "arXiv",
    primaryClass = "nucl-th",
    month = "6",
    year = "2020"
}

@article{Bally:2021qys,
    author = "Bally, Benjamin and Bender, Michael and Giacalone, Giuliano and Som\`a, Vittorio",
    title = "{Evidence of the triaxial structure of $\boldsymbol{^{129}}$Xe at the Large Hadron Collider}",
    eprint = "2108.09578",
    archivePrefix = "arXiv",
    primaryClass = "nucl-th",
    month = "8",
    year = "2021"
}

@article{Giacalone:2021udy,
    author = "Giacalone, Giuliano and Jia, Jiangyong and Zhang, Chunjian",
    title = "{Impact of Nuclear Deformation on Relativistic Heavy-Ion Collisions: Assessing Consistency in Nuclear Physics across Energy Scales}",
    eprint = "2105.01638",
    archivePrefix = "arXiv",
    primaryClass = "nucl-th",
    doi = "10.1103/PhysRevLett.127.242301",
    journal = "Phys. Rev. Lett.",
    volume = "127",
    number = "24",
    pages = "242301",
    year = "2021"
}

@Book{bohr,
    editor = {Bohr, Aage and Mottelson, Ben R},
    title = "{Nuclear Structure}",
    doi = " https://doi.org/10.1142/3530",
    isbn = "978-981-238-660-1",
    publisher = "World Scientific",
    year = "1998"
}

@article{Giacalone:2021clp,
    author = {Giacalone, Giuliano and Schenke, Bj\"orn and Shen, Chun},
    title = "{Constraining the Nucleon Size with Relativistic Nuclear Collisions}",
    eprint = "2111.02908",
    archivePrefix = "arXiv",
    primaryClass = "nucl-th",
    doi = "10.1103/PhysRevLett.128.042301",
    journal = "Phys. Rev. Lett.",
    volume = "128",
    number = "4",
    pages = "042301",
    year = "2022"
}

@article{Jia:2021qyu,
    author = "Jia, Jiangyong",
    title = "{Probing triaxial deformation of atomic nuclei in high-energy heavy ion collisions}",
    eprint = "2109.00604",
    archivePrefix = "arXiv",
    primaryClass = "nucl-th",
    doi = "10.1103/PhysRevC.105.044905",
    journal = "Phys. Rev. C",
    volume = "105",
    number = "4",
    pages = "044905",
    year = "2022"
}

@article{ATLAS:2022dov,
    author = "{ATLAS Collaboration}",
    xcollaboration = "ATLAS",
    title = "{Correlations between flow and transverse momentum in Xe+Xe and Pb+Pb collisions at the LHC with the ATLAS detector: a probe of the heavy-ion initial state and nuclear deformation}",
    eprint = "2205.00039",
    archivePrefix = "arXiv",
    primaryClass = "nucl-ex",
    reportNumber = "CERN-EP-2022-052",
    month = "4",
    year = "2022"
}

@article{Zhang:2021kxj,
    author = "Zhang, Chunjian and Jia, Jiangyong",
    title = "{Evidence of Quadrupole and Octupole Deformations in Zr96+Zr96 and Ru96+Ru96 Collisions at Ultrarelativistic Energies}",
    eprint = "2109.01631",
    archivePrefix = "arXiv",
    primaryClass = "nucl-th",
    doi = "10.1103/PhysRevLett.128.022301",
    journal = "Phys. Rev. Lett.",
    volume = "128",
    number = "2",
    pages = "022301",
    year = "2022"
}

\end{document}